\documentclass[aps,prx,twocolumn,floatfix]{revtex4}
\usepackage{latexsym}
\usepackage{subfigure}
\usepackage{epsfig}
\usepackage{graphicx}
\usepackage{dcolumn}
\usepackage{bm}
\usepackage{epstopdf}
\usepackage{natbib}
\usepackage{bm}
\usepackage{color}

\usepackage{subfigure}
\usepackage{hyperref}

\begin{document}
\title{Bulk Fermi Surfaces of the Dirac Type-II Semimetallic Candidates $M$Al$_3$ \\
(where \emph{M} = V, Nb and Ta)}

\author{K. -W. Chen,$^{1,2}$}\email{kchen@magnet.fsu.edu}
\affiliation{$^1$National High Magnetic Field Laboratory, Florida State University, Tallahassee, Florida 32310, USA}
\affiliation{$^2$Department of Physics, Florida State University, Tallahassee, Florida 32306, USA}
\author{X. Lian,$^{1,2}$ Y. Lai,$^{1,2}$ N. Aryal,$^{1,2}$ Y. -C. Chiu,$^{1,2}$ W. Lan,$^{1,2}$\\
 D. Graf,$^{1}$ E. Manousakis,$^{1,2}$ R. E. Baumbach,$^{1,2}$}
\affiliation{$^1$National High Magnetic Field Laboratory, Florida State University, Tallahassee, Florida 32310, USA}
\affiliation{$^2$Department of Physics, Florida State University, Tallahassee, Florida 32306, USA}
\author{L. Balicas$^{1,2}$}\email{balicas@magnet.fsu.edu}
\affiliation{$^1$National High Magnetic Field Laboratory, Florida State University, Tallahassee, Florida 32310, USA}
\affiliation{$^2$Department of Physics, Florida State University, Tallahassee, Florida 32306, USA}
\date{\today}

\begin{abstract}
 We report a de Haas-van Alphen (dHvA) effect study on the Dirac type-II semimetallic candidates \emph{M}Al$_3$ (where, \emph{M} = V, Nb and Ta). The angular-dependence of their Fermi surface (FS) cross-sectional areas reveals a remarkably good agreement with first-principle calculations. Therefore, dHvA supports the existence of tilted Dirac cones with Dirac type-II nodes located at 100, 230 and 250 meV above the Fermi level $\varepsilon_F$ for VAl$_3$, NbAl$_3$ and TaAl$_3$ respectively, in agreement with the prediction of broken Lorentz invariance in these compounds. However, for all three compounds we find that the cyclotron orbits on their FSs, including an orbit nearly enclosing the Dirac type-II node, yield trivial Berry phases. We explain this \emph{via} an analysis of the Berry phase where the position of this orbit, relative to the Dirac node, is adjusted within the error implied by the small disagreement between our calculations and the experiments. We suggest that a very small amount of doping could displace $\varepsilon_F$ to produce topologically non-trivial orbits encircling their Dirac node(s).
\end{abstract}

\maketitle

Condensed-matter systems provide accessible platforms for the discovery of quasiparticles with properties akin to particles predicted by high-energy physics.
 Dirac type-I compounds such as Cd$_3$As$_2$ \cite{Liu14, Wang13} or Na$_3$Bi, \cite{Wang12, ZKLiu14}) and Weyl type-I systems like (Ta,Nb)(P,As) \cite{Weng15, Lv15, Xu15} were recently discovered and are garnering a lot of attention. According to Ref. \cite{bernevig_science_2016}, solid state systems would even offer the potential of finding fermionic excitations which have no analog in high-energy physics such as three-component fermions.\cite{ding_nature} Very recently, so-called type-II Dirac/Weyl semimetals, which violate Lorentz-symmetry, were discovered and are being intensively studied. In type-II Dirac/Weyl semimetals, the crossings between energy bands remain protected but the spectra of the  Dirac/Weyl cones are strongly tilted due to an additional momentum dependent term, thus they break the Lorentz invariance. These Weyl/Dirac type-II nodes become singular points connecting electron and hole pockets in the spectral function. The associated quasiparticles can be observed in condensed matter systems but the analog particles are absent in Lorentz invariant high energy physics. Type-II Dirac/Weyl semimetallic systems were proposed to display unique properties, such as Klein tunneling in momentum space,\cite{Brien16} orientation-dependent chiral anomaly,\cite{Udagawa16} and a modified anomalous Hall conductivity.\cite{Zyuzin16}

A number of Weyl type-II semimetals were already experimentally studied, including MoTe$_2$,\cite{Deng16, Kam16, Wang16} WTe$_2$,\cite{Bruno16,Huang16,Peng16} Ta$_3$S$_2$,\cite{Chang16} LaAlGe,\cite{Xu17} and TaIrTe$_4$.\cite{Koepernik16, Khim16} And although several Dirac type-II compounds were also reported, e.g. VAl$_3$,\cite{Chang17,Ge} YPd$_2$Sn,\cite{Guo17} KMgBi,\cite{Le17} and (Pt,Pd)Te$_2$,\cite{Yan17,Noh17,Zhang17} only the last two compounds were studied experimentally \emph{via} angle-resolved photoemission spectroscopy (ARPES) and quantum oscillatory phenomena.\cite{Yan17,Noh17,Fei} Therefore, Dirac type-II systems remain to be unambiguously identified and characterized experimentally. For instance, and in addition to the aforementioned predictions, Dirac type-II semimetals have been predicted to become Weyl type-II systems or topological crystalline insulators when time-reversal or inversion symmetries are broken.\cite{Chang17,Ge} However, exposing the unique transport properties of such compounds is a difficult task due to the distance of the Dirac type-II nodes with respect to the Fermi level which, in addition, is crossed by topologically trivial and non-trivial bands.

Here, we report the synthesis of the chemical analogues (V, Nb, Ta)Al$_3$ which were predicted to display Dirac type-II nodes.\cite{Chang17}
The topography of their Fermi surfaces, revealed through the de Haas van Alphen (dHvA) effect, are found to display remarkably good agreement with band structure calculations.
Therefore, our experimental study indicates that these compounds would break Lorentz invariance.\cite{Chang17} The Dirac type-II nodes in VAl$_3$ and NbAl$_3$ are found to be relatively close to Fermi level, i.e. at respectively $\approx$ 100 meV and $\approx$ 230 meV above it, making these compounds promising candidates for tuning the Fermi level (e.g. \emph{via} chemical substitution) towards the linearly dispersive regions of their bands. In fact, we find that one of the observed cyclotron orbits nearly encloses the Dirac type-II node(s), and although it yields a topologically trivial Berry-phase, a small amount of doping or displacement of $\varepsilon_F$ could lead to a topologically non-trivial orbit.
\begin{figure*}[htb]
    \begin{center}
 \includegraphics[width=13 cm]{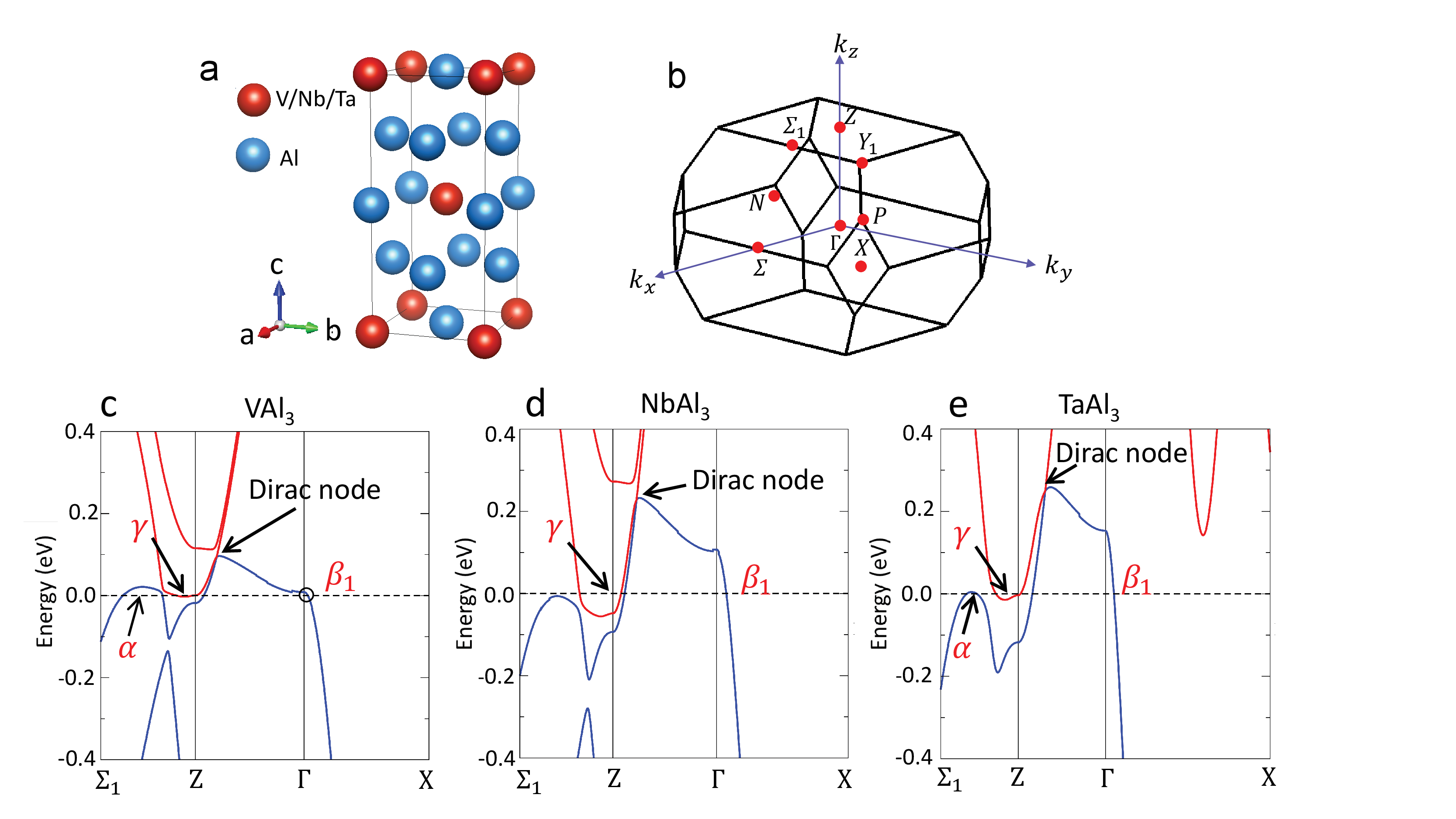}
	\caption{(a) Crystal structure of the \emph{M}Al$_3$ family. Red spheres represents V, Nb, or T while the blue ones depict Al atoms. (b) Brillouin zone and high symmetry points.
(c)(d)(e) The calculated band structures for all three compounds. Dirac type-II nodes along the Z$-\Gamma$ lines, and the Fermi surface cross-sectional areas associated with the  $\alpha$, $\beta_1$, and $\gamma$ dHvA-orbits are indicated by arrows.}
	\label{fig:xray}
    \end{center}
\end{figure*}

Details concerning single-crystal growth can be found in the Supplemental Information (SI) file.\cite{supplemental} Supplemental Figs. S1, S2, S3 and S4 provide X-ray diffraction, resistivity, a discussion on the role of Al inclusions,\cite{li} and values for the extracted mobilities, respectively.\cite{supplemental} Density functional theory calculations were performed with the Wien2K~\cite{wien2k} implementation of Density Functional Theory (DFT) the Perdew-Burke-Ernzerhof parametrization of the generalized gradient approximation (GGA-PBE) ~\cite{PBE}. The angular dependency of the  Fermi-surface cross-sectional area were computed through SKEAF.\cite{SKEAF} The validity of these results were verified through the Quantum Expresso implementation of DFT obtaining very similar results.\cite{QE1,QE2}

Here, we report a study on the electronic structure at the Fermi level of the \emph{M}Al$_3$ family through the dHvA-effect superimposed onto torque magnetometry.
The \emph{M}Al$_3$ compounds crystalize in a body-centered tetragonal Bravais lattice belonging to the space group $I$4/$mmm$ (No. 139) as shown in Fig.~\ref{fig:xray}(a). Lattice constants determined through X-ray diffraction are given in the SI.\cite{supplemental} According to DFT, there are two tilted Dirac cones along the Z$-\Gamma-$Z line within the first Brillouin zone (see, Fig.~\ref{fig:xray}(c)), with the Dirac node located at the touching point between the electron and hole cones. As seen through Figs.~\ref{fig:xray}(c) to \ref{fig:xray}(e), the crossing of the hole bands with the Fermi level produces two hole-like Fermi surface sheets, i.e. the $\alpha$ and the $\beta$ orbits while the electron bands lead to the $\gamma$ pockets. The proximity of the $\beta$ and $\gamma$ orbits to the Dirac type-II nodes would make them prime candidates for carriers displaying non-trivial Berry phases.
In contrast, we would expect the $\alpha$-pocket along the $\Sigma_1-$Z line, which is absent in NbAl$_3$, to yield topologically trivial orbits. As we show below, the electron pocket displays a ``helix" like shape while the hole one yields a ``dumbbell" like sheet which supports two orbits: neck (or the $\beta_1$-orbit) and belly (or the $\beta_2$-orbit).

\begin{figure}[htb]
    \begin{center}
 \includegraphics[width = 8.6 cm]{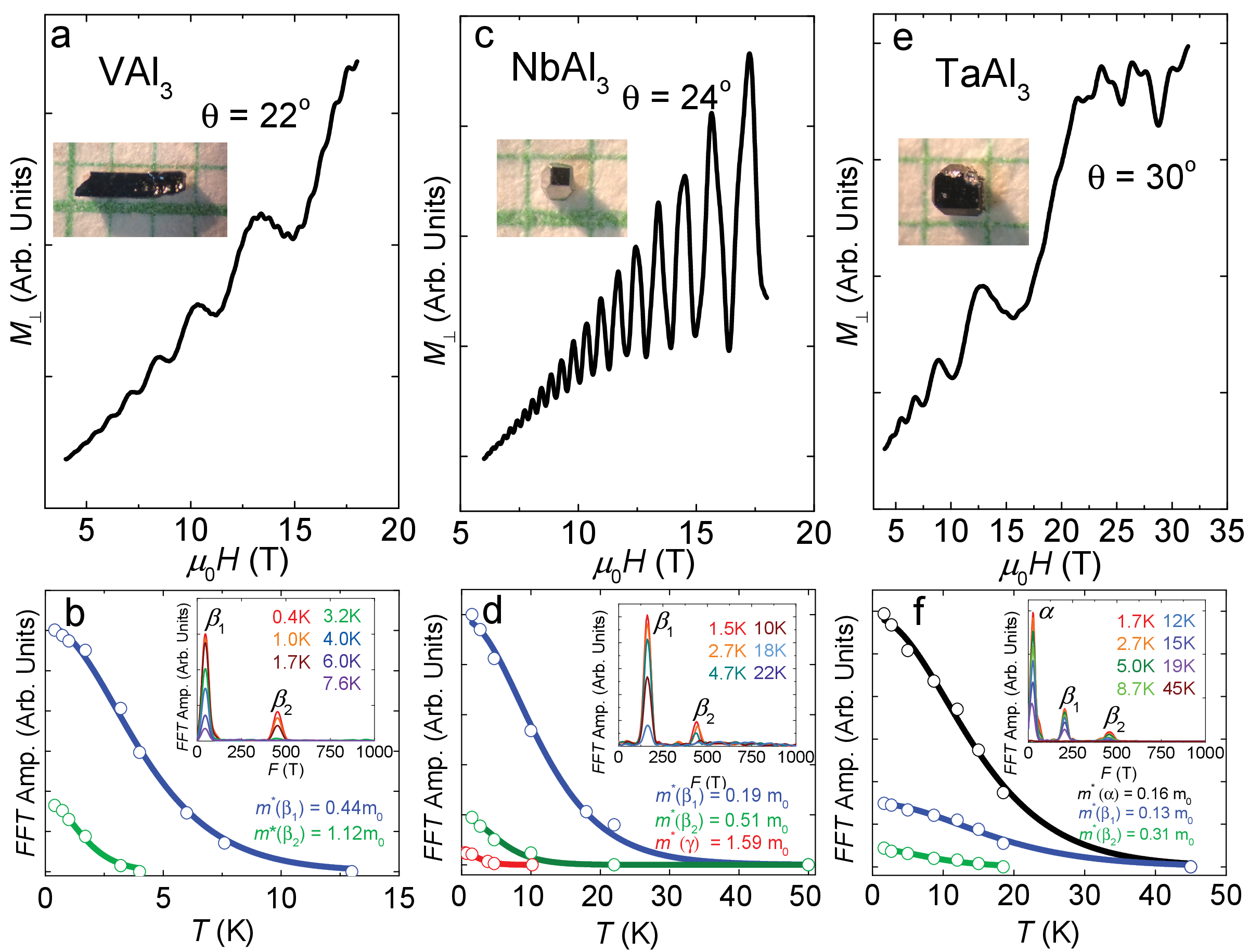}
	\caption{Transverse component of the magnetization $M_{\perp}$ as a function of magnetic field $\mu_0H$ for (a) VAl$_3$, (c) NbAl$_3$, and (e) TaAl$_3$, respectively. These traces were collected at angles  $\theta = 22^{\circ}$, 24$^{\circ}$, and 30$^{\circ}$ respectively, between the external field and the \emph{c}-axis. These traces were measured at a temperature $T = 0.4$ K. (b), (d), (f) Amplitude of the main peaks observed in the Fourier spectra of the oscillatory signal extracted from each compound as a function of the temperature. Insets: Fast Fourier transforms of the dHvA signal extracted from each compound at several temperatures.}
	\label{fig:raw}
    \end{center}
\end{figure}

The torque signal $\tau(\mu_0H)$ can be expressed in terms of the component of the magnetization perpendicular to the external field $\mu_0H$:
 $M_{\perp}$ = $\tau$/$V\mu_0H$ where $V$ is the volume of the sample.  
From the Onsager relation, the frequencies $F$ of the oscillatory signal are proportional to the extremal cross-sectional areas $A$ of a given Fermi surface sheet:
\begin{eqnarray}
F = \frac {\hbar}{2\pi e}A
\end{eqnarray}
 where $\hbar$ is the reduced Planck constant and $e$ the electrical charge. Figures~\ref{fig:raw}(a), ~\ref{fig:raw}(c), and ~\ref{fig:raw}(e) display $M_{\bot}$ as a function of $\mu_0H$, collected at a temperature $T = 0.4$ K, for the V, Nb and Ta compounds respectively, where a superimposed oscillatory signal, or the dHvA effect, can be observed. These traces were collected at angle $\theta = 22^{\circ}$ (for the V compound), $24^{\rm{o}}$ (Nb) and $30^{\circ}$ (Ta) respectively, where $\theta$ = 0$^{\circ}$ and 90$^{\circ}$ correspond to fields along the \emph{c-} and the \emph{a-}axes, respectively. The anomaly observed in TaAl$_3$ near $\mu_0H \simeq 20$ T is most likely an indication for the quantum limit associated to the $\alpha$-orbit. Figures ~\ref{fig:raw}(b), ~\ref{fig:raw}(d), and ~\ref{fig:raw}(f), display the temperature dependence of the main peaks/frequencies observed in the FFT spectra of the oscillatory signal extracted for each compound, see insets. The dHvA signals were obtained after fitting the background of $M_{\bot}$ to a polynomial and its subsequent subtraction. By fitting the amplitude of the FFT peaks as a function of the temperature to the thermal damping term in the Lifshitz-Kosevich formalism \cite{shoenberg} (see, Eq. (2)) one can extract the carrier effective masses $m^{\ast}$, see Figs. \ref{fig:raw}(b), \ref{fig:raw}(d), and \ref{fig:raw}(f). For the neck $F_{\beta_{1}} = 35$ T and belly $F_{\beta_{2}}=410 $ T orbits on the ``dumbbell" hole-pocket of VAl$_3$, we extracted $m^{\ast}_{\beta_1}$ = 0.44 $m_0$ and $m^{\ast}_{\beta_2}$ = 1.12 $m_0$, respectively. Here, $m_0$ is the free-electron mass. As we show below, these orbits were identified after a detailed comparison with the DFT calculations. For NbAl$_3$ we extracted $F_{\beta_1} = 163$ T with $m^{\ast}_{\beta_1}$ = 0.19 $m_0$ and $F_{\beta_2} = 440$ T with corresponding $m^{\ast}_{\beta_2}$ = 0.51 $m_0$. For TaAl$_3$ we identified $F_{\beta_1}$ = 174 T with $m^{\ast}_{\beta_1} = 0.13$ $m_0$ and $F_{\beta_2} = 423$ T with $m^{\ast}_{\beta_2} = 0.31$ $m_0$. We are also able to detect the large electron $\gamma$ pocket of NbAl$_3$ through magnetoresistivity or Shubnikov-de Haas (SdH) measurements under fields up to $\mu_0H = 31$ T yielding $F_{\gamma} \simeq 3300$ T with $m^{\ast}_{\gamma} = 1.59$ $m_0$. These $m^{\ast}$ values are not particularly light which indicates that these orbits are not very close to the linearly dispersive region of the bands.
\begin{figure}[htb]
    \begin{center}
 \includegraphics[width = 8.6 cm]{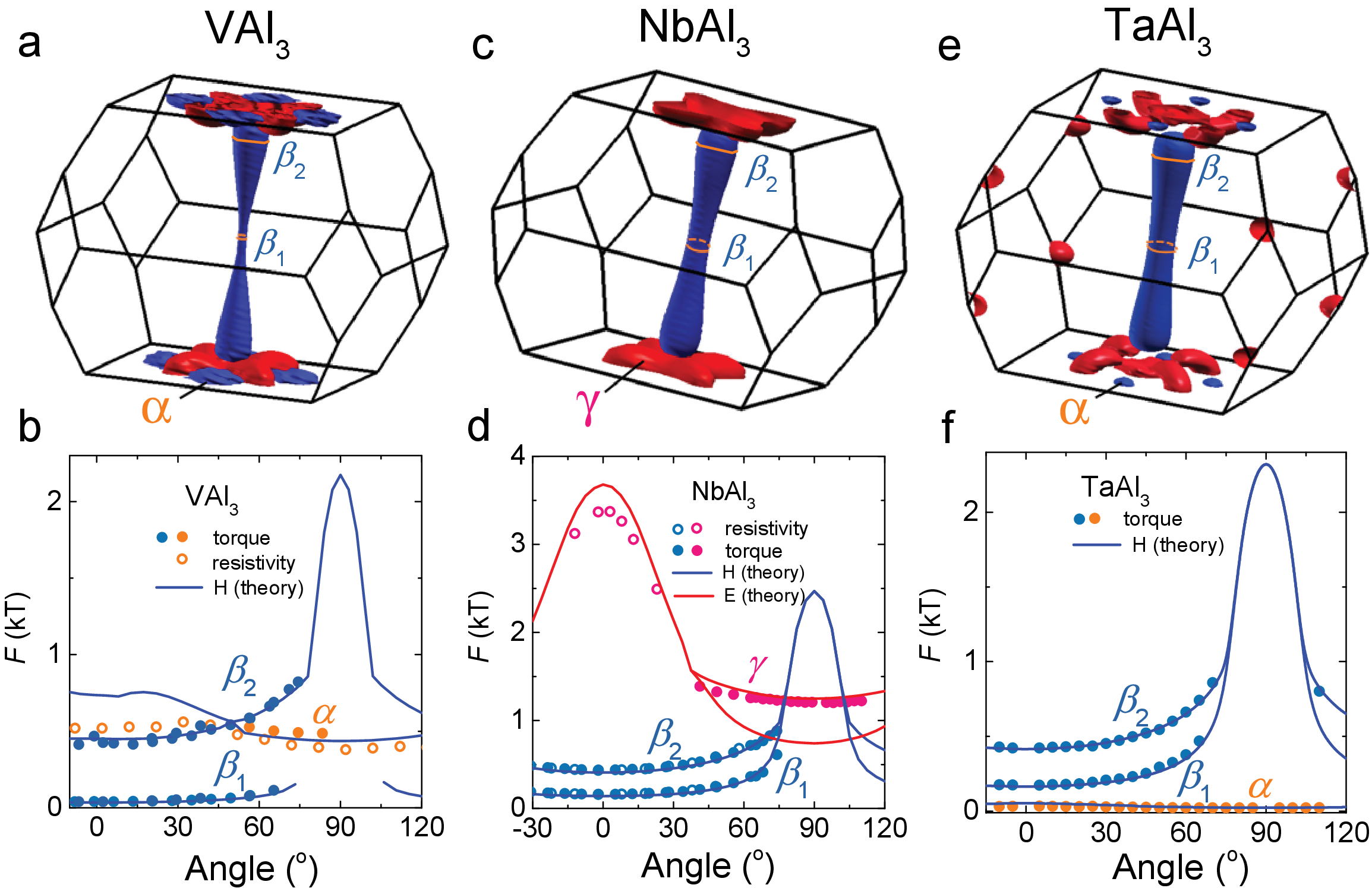}
	\caption{Fermi surfaces for (a) VAl$_3$ (c) NbAl$_3$ (e) TaAl$_3$ respectively. Hole- and electron-like pockets are depicted in blue and in red, respectively. $\beta_1$ and $\beta_2$ orbits match the frequencies calculated for the ``neck" and the ``belly" cross-sectional areas of the dumbbell like pockets. $\alpha$ orbit can be associated to a hole-like ellipsoid of topologically trivial character. $\gamma$ orbit can be associated with the large ``helix" like electron pocket. (b), (d), (f) Cyclotron frequencies $F$ as functions of the angle $\theta$ relative to crystallographic \emph{c}-axis. Open and closed symbols depict SdH and dHvA data, respectively. Solid lines depict the angular dependence of the FS extremal cross-sectional areas predicted theoretically.}
	\label{fig:DFT}
    \end{center}
\end{figure}
 \begin{figure*}[htb]
    \begin{center}
 \includegraphics[width= 13 cm]{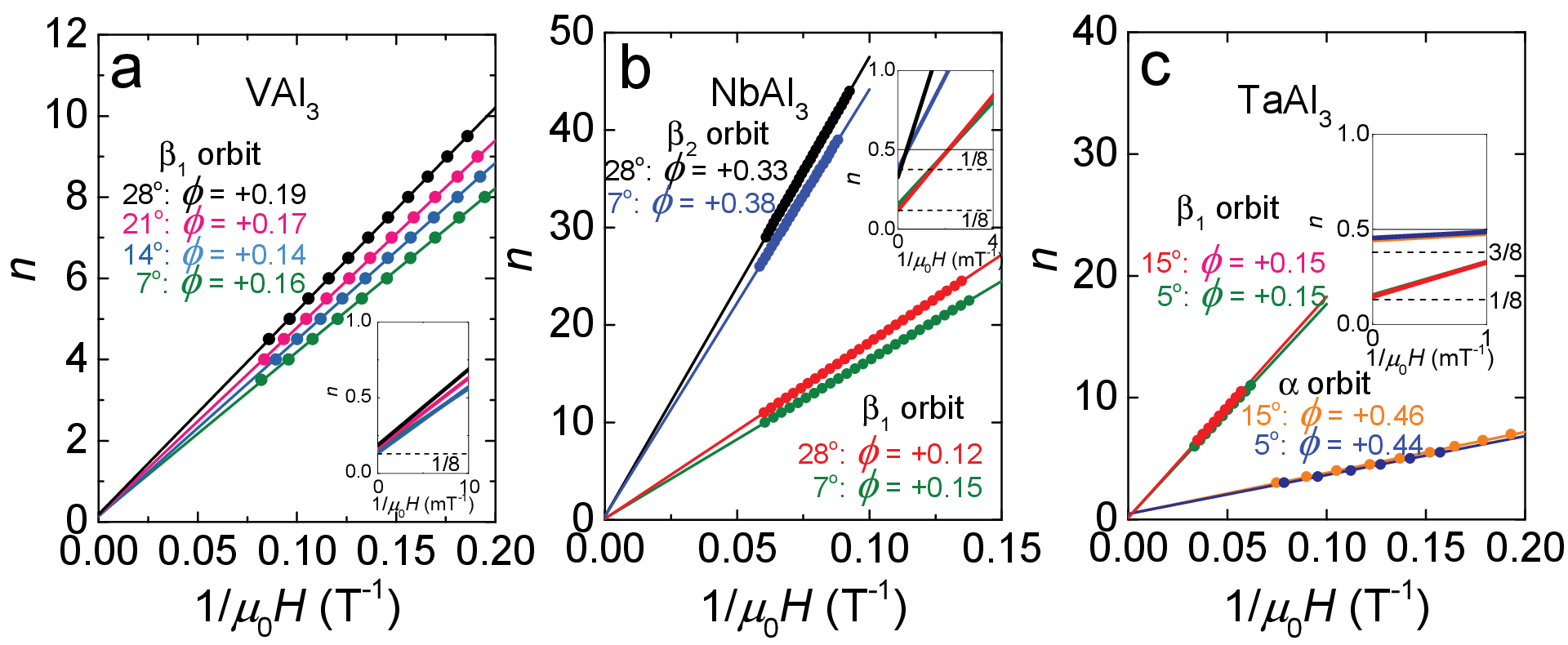}
	\caption{(a), (b), (c) Landau level index $n$ as a function of $(\mu_0H)^{-1}$ for all three compounds and for several angles $\theta$ between the external field and the \emph{c}-axis. Here, the $n$ indexes were assigned to the peaks in the longitudinal resistivity $\Delta\rho_{xx}$ or in the magnetic susceptibility $\Delta\chi$ while $n+1/2$ were assigned to the minima. The intercept yields the phase $\phi_B$ which acquires values of $\sim 1/8$ for $\beta_1$ orbit and of $\sim 3/8$ for the $\alpha$ and $\beta_2$ orbits.}
	\label{fig:Berry}
    \end{center}
\end{figure*}

In order to study the geometry of the Fermi surface, we measured the angular dependence of the dHvA oscillations. A comparison between the DFT calculations and the angular dependence of the dHvA frequencies shown in Figs. 3(a) to 3(f), indicates that the $\beta_1$ and $\beta_2$ orbits of the \emph{M}Al$_3$ compounds correspond to the minimum and maximum cross-sectional areas of the ``dumbbell" hole-pockets (in blue, around the $\Gamma-$points). In Figures 3(b), 3(d), and 3(f), markers correspond to the experimental points while lines depict the angular dependence of the FS cross-sectional areas according to DFT. The topologically trivial $\alpha$ orbit corresponds to ellipsoids (in blue) extending along the $Z-\Sigma_1$ lines in VAl$_3$ and in TaAl$_3$ which are absent in NbAl$_3$. Finally, the $\gamma$-orbit corresponds to the electron-like ``helix" sheet, depicted in red around the $Z$-point. As seen through Figs. 3(b), 3(d), and 3(f), the experimentally obtained values for the $\beta_1$ and $\beta_2$ orbits agree very well with the calculated ones but become unmeasurable for $\theta > 75^{\circ}$. This suggests either very anisotropic effective masses or, most likely, anisotropic scattering rates.
The $\alpha$ orbit follows the theoretical predictions with some deviations which can be partially attributed to sample misalignment. As for the $\gamma$-orbit, it was observable only in NbAl$_3$, hence we chose not to plot the respective theoretical traces for VAl$_3$ and TaAl$_3$ in Figs. 3(b) and 3(f). Its non-observation is likely the result of large cross-sectional areas (or frequencies) combined with heavier effective masses leading to low mobilities. The very good agreement between the experimentally determined and the calculated FS cross-sectional areas supports the existence of a Dirac Type-II node located at $\simeq 100$, $\simeq 230$, and $250$ meV above the Fermi level in VAl$_3$, NbAl$_3$, and TaAl$_3$, respectively. Notice that our calculations do not yield the same FS topography as the ones reported in Refs.\cite{Chang17,Ge}. For instance, in their calculations the dumbbell like hole-pocket would have a quite different geometry. The exact position of the Dirac nodes relative to $\varepsilon_F$ also differs between our calculations and those in Refs.\cite{Chang17,Ge}.

Now, we address the topological character of the observed orbits; the Lifshitz-Kosevich formalism describing the field and the temperature-dependence of the dHvA oscillations is given by:\cite{shoenberg}
\begin{eqnarray}
M_{\parallel} = - AB^{1/2}R_TR_DR_S \sin \left[ 2\pi \left(\frac{F}{B}-\gamma+\delta \right)\right]
\end{eqnarray}
where $\gamma = 1/2$ for a parabolic band and $ = 0$ for a linear one. The phase shift $\delta$ is determined by the dimensionality of the Fermi surface taking values $\delta=\pm 1/8$ for minima and maxima cross-sectional areas of a three-dimensional Fermi surface, respectively. $R_{T} = X/\sinh(X$) is the thermal damping factor, where $X = 2\pi^{2} k_Bm^{*}T/e\hbar B$, $R_{D} = \exp(-2\pi^{2}k_Bm^{*}T_D/e\hbar B)$ is the Dingle damping factor, where $m^{*}$ is the effective mass and $T_D = \hbar/2\pi k_B\tau_q$ from which one can evaluate the quasiparticle lifetime $\tau_q$. $R_{s} = \cos(\pi /2 g m^{\ast}/m_0)$ is the spin damping factor, where $g$ is the Land\`{e} $g$-factor. In order to extract the correct phase of the dHvA oscillations, we make use of the magnetic susceptibility $\Delta \chi = d(\Delta M)/d(\mu_0H)$:
\begin{eqnarray}
\Delta\chi\sim \text{sign} R_{S}\cos\left[2\pi\left(\frac{F}{B}-\gamma+\delta\right)\right]
\end{eqnarray}

The phase of the oscillations is given by $\phi = -\left( \gamma-\delta\right)$ where $\gamma= (1/2 - \phi_B/2\pi)$, with $\phi_B$ being the Berry phase, and $\delta =0$ or $\pm 1/8$ for  two- and three-dimensional (3D) FSs, respectively. For trivial 3D bands  $\phi_B=0$, hence one expects $\phi= 1/2 \mp 1/8 = 3/8$ or 5/8, for maximum and minimum cross-sectional areas, respectively. Notice that the value of $\phi$ can be affected by the sign of the spin damping factor $R_S$ and this has to be carefully considered when extracting the Berry phase. To experimentally extract the phase, we assign integer Landau level indices $n$ to the peaks in $\Delta\chi$ (maxima in the density of states \cite{wang}) and $n+1/2$ to the valleys. The phase $\phi$ can be extracted from the intercept of the extrapolation of $n$ as function of $(\mu_0H)^{-1}$; Landau fan diagrams shown in Figs. Fig.~\ref{fig:Berry}(a),
Fig.~\ref{fig:Berry}(b), and Fig.~\ref{fig:Berry}(c), for VAl$_3$, NbAl$_3$, and TaAl$_3$, respectively. For all the three compounds, we obtain $\phi \sim +1/8$ for the $\beta_1$ orbit which encircles the $\Gamma$-point in the FBZ which, according to DFT, corresponds to a minimal cross-sectional area of a trivial parabolic band. Hence, this anomalous 1/8 value can be understood as $-1/2+5/8$, where the -1/2 term is attributable to a ``minus" sign provided by the spin dephasing term $R_s$.
The $\beta_2$ orbit is the one encircling the Dirac node. However, it is difficult to extract $\phi$ for the $\beta_2$-orbit of both the VAl$_3$ and TaAl$_3$ compounds given that their higher frequencies are superimposed onto those of the $\alpha$ and $\beta_1$ orbits. Fortunately, we were able to extract $\phi \sim  3/8$ for the $\beta_2$ orbit of NbAl$_3$ through torque measurements. As for the $\alpha$-orbit associated with the trivial parabolic band of TaAl$_3$, we find $\phi \simeq  0.44  \gtrsim 3/8$, see Fig.~\ref{fig:Berry}(c). This value was confirmed by SQUID magnetometometry measurements. See, supplemental Figs. S5, S6, S7 and S8 for band pass filter and phase analysis of the dHvA signal, Berry-phase analysis based on resistivity, based on SQUID magnetometry, and analysis of the spin de-phasing term for VAl$_3$, NbAl$_3$, and TaAl$_3$, respectively.\cite{supplemental} In the same Ref. \onlinecite{supplemental} we provide a calculation of the Berry-phase based upon the model of Ref. \cite{Chang17}. We show that within the error implied by very small displacements in $\varepsilon_F$ introduced into the DFT calculations to match the experimental results, the $\beta_2$ orbit could not enclose the Dirac node(s) yielding a Berry phase $< \pi/2$.

In summary, we unveiled the Fermi surfaces of the \emph{M}Al$_3$ family through quantum oscillations measurements combined with band structure calculations. Among all three compounds, VAl$_3$ displays the closest Dirac type-II node with respect to its Fermi level ($\sim$ 100 meV). The extracted Berry phases for all of the measured Fermi surfaces are consistent with time-reversal-symmetric systems displaying discrete and topologically trivial values. Although the $\beta_2$ orbit of NbAl$_3$ nearly encloses the Dirac node, it also leads to the observation of a trivial Berry-phase due to its exact position in the $k_z$ plane relative to the $k_z^{\text{DP}}$ position of the Dirac type-II point. As discussed in Ref. \cite{wang}, the Berry phase can quickly become trivial as the Fermi level is displaced away from the Dirac node(s). However, our calculations indicate that the Dirac node is displaced towards the Fermi level as the ionic size of the transition metal decreases, indicating a role for chemical substitution. One needs only a very small displacement in $\varepsilon_F$ to stabilize a topologically non-trivial $\beta_2$ orbit enclosing the Dirac node. Notice that this approach might also contribute to stabilize bulk superconductivity \cite{NbAl3} or a Weyl type-II state if one chose magnetic dopants.\cite{Chang17}

The authors thank H. -Z. Lu at South University of Science and Technology of China for informative discussions and M. Khan at Louisiana State University for help with our measurements. This work was supported by DOE-BES through award DE-SC0002613. K.W.C. was partially supported by the NHMFL-UCGP program. The NHMFL is supported by NSF through NSF-DMR-1157490 and the State of Florida. Correspondence and requests for materials should be addressed to K.W.C or to L.B.

\end{document}